\documentclass{svjour3}

\usepackage{cite}

\begin{document}

\title{Non-Hermitian Hamiltonians and similarity transformations}
\author{Francisco M. Fern\'{a}ndez}

\institute{INIFTA (UNLP, CCT La Plata-CONICET), Divisi\'on
Qu\'imica Te\'orica, Blvd. 113 S/N, Sucursal 4, Casilla de Correo
16, 1900 La Plata, Argentina }

\date{Received: date / Accepted: date}
\maketitle

\begin{abstract}
We show that similarity (or equivalent) transformations enable one
to construct non-Hermitian operators with real spectrum. In this
way we can also prove and generalize the results obtained by other
authors by means of a gauge-like transformation and its
generalization. Such similarity transformations also reveal the
connection with pseudo-Hermiticity in a simple and straightforward
way. In addition to it we consider the positive and negative
eigenvalues of a three-parameter non-Hermitian oscillator.

\end{abstract}

\section{Introduction}

\label{sec:inrto}

In the last years there has been great interest in the mathematical
properties of non-Hermitian Hamiltonians, which was mainly aroused by the
conjecture that the non-Hermitian Hamiltonians with real spectra studied so
far\cite{A95, DP98, FGRZ98} exhibited PT symmetry\cite{BB98}. There is a
vast literature on non-Hermitian Hamiltonians, some of which is reviewed
elsewhere\cite{B07}. Later Mostafazadeh\cite{M02a,M02b,M02c} showed that
every Hamiltonian with a real spectrum is pseudo-Hermitian and that all the
PT-symmetric Hamiltonians studied in the literature exhibited such property.
On the other hand, the so-called space-time symmetry did not prove to be so
robust in producing non-Hermitian operators with real spectra\cite
{FG14a,FG14b,AFG14b,AFG14c}.

Some time ago Ahmed\cite{A02} derived a family of one-dimensional
non-Hermitian Hamiltonians with real spectrum by means of a \textit{%
gauge-like} transformation. He argued that the eigenfunctions of the
resulting PT-symmetric Hamiltonian did not satisfy the PT-orthogonality
condition. Recently, Rath and Mallick\cite{RM15} put forward a
generalization of the gauge-like transformation that involves both the
coordinate and momentum operators and leads to a non-Hermitian Hamiltonian
that appears to be isospectral with the harmonic oscillator.

The purpose of this paper is to discuss the gauge-like transformation in a
more general and rigorous setting. In Section~\ref{sec:similarity} we
outline the main ideas of the similarity (or equivalent) transformation
between a non-Hermitian and a Hermitian Hamiltonian. In Section~\ref
{sec:gauge} we discuss the gauge-like transformation introduced by Ahmed and
in Section~\ref{sec:trans_x_p} the somewhat more general transformation
proposed by Rath and Mallick. In Section~\ref{sec:gen_x,p_transf} we show
how to generalize the latter. In Section\ref{sec:pos_neg_eigen} we discuss a
somewhat more general three-parameter non-Hermitian oscillator and obtain
its eigenvalues and eigenvectors in a somewhat different way. Finally, in
Section~\ref{sec:conclusions} we summarize the main results and draw
conclusions.

\section{Similarity or equivalent transformation}

\label{sec:similarity}

Let $H$ be a Hermitian operator with a discrete spectrum
\begin{equation}
H\psi _{n}=E_{n}\psi _{n},  \label{eq:Hpsi_n}
\end{equation}
and a complete set of eigenvectors
\begin{equation}
\sum_{n}\left| \psi _{n}\right\rangle \left\langle \psi _{n}\right|
=I,\;\left\langle \psi _{m}\right. \left| \psi _{n}\right\rangle =\delta
_{mn},  \label{eq:completeness}
\end{equation}
where $I$ is the identity operator. Its spectral decomposition reads
\begin{equation}
H=\sum_{n}E_{n}\left| \psi _{n}\right\rangle \left\langle \psi _{n}\right| .
\label{eq:spectral_desc}
\end{equation}

For every linear invertible operator $U$ the similarity transformation
\begin{equation}
\tilde{H}=UHU^{-1},  \label{eq:H'}
\end{equation}
yields a new operator $\tilde{H}$ that is not Hermitian unless $%
U^{-1}=U^{\dagger }$. We say that $H$ and $\tilde{H}$ are equivalent or
similar. The transformed vectors
\begin{equation}
\left| \varphi _{n}\right\rangle =U\left| \psi _{n}\right\rangle ,
\label{eq:Upsi_n}
\end{equation}
are eigenvectors of $\tilde{H}$
\begin{equation}
\tilde{H}\left| \varphi _{n}\right\rangle =UHU^{-1}U\left| \psi
_{n}\right\rangle =E_{n}\left| \varphi _{n}\right\rangle ,
\label{eq:H'Upsi_n}
\end{equation}
whereas
\begin{equation}
\left| \Phi _{n}\right\rangle =\left( U^{-1}\right) ^{\dagger }\left| \psi
_{n}\right\rangle ,  \label{eq:Phi_n}
\end{equation}
are eigenvectors of the adjoint operator $\tilde{H}^{\dagger }$
\begin{equation}
\tilde{H}^{\dagger }\left| \Phi _{n}\right\rangle =\left( U^{-1}\right)
^{\dagger }HU^{\dagger }\left( U^{-1}\right) ^{\dagger }\left| \psi
_{n}\right\rangle =E_{n}\left| \Phi _{n}\right\rangle .
\end{equation}
Both sets of vectors form a biorthonormal basis
\begin{equation}
\left\langle \Phi _{m}\right. \left| \varphi _{n}\right\rangle =\left\langle
\psi _{m}\right. \left| \psi _{n}\right\rangle =\delta _{mn},
\end{equation}
that enables us to write
\begin{equation}
\tilde{H}=\sum_{n}E_{n}U\left| \psi _{n}\right\rangle \left\langle \psi
_{n}\right| U^{-1}=\sum_{n}E_{n}\left| \varphi _{n}\right\rangle
\left\langle \Phi _{n}\right| .
\end{equation}

The basis set $\{\left| \varphi _{n}\right\rangle \}$ is orthonormal with
the metric given by $\left( U^{-1}\right) ^{\dagger }U^{-1}$:
\begin{equation}
\left\langle \psi _{m}\right. \left| \psi _{n}\right\rangle =\left\langle
\varphi _{m}\right| \left( U^{-1}\right) ^{\dagger }U^{-1}\left| \varphi
_{n}\right\rangle =\delta _{mn}.  \label{eq:metric}
\end{equation}
On the other hand, the standard inner product
\begin{equation}
\left\langle \varphi _{m}\right. \left| \varphi _{n}\right\rangle
=\left\langle \psi _{m}\right| U^{\dagger }U\left| \psi _{n}\right\rangle ,
\end{equation}
is not necessarily finite.

It follows from (\ref{eq:H'}) that
\begin{equation}
\tilde{H}^{\dagger }=\left( U^{-1}\right) ^{\dagger }HU^{\dagger }=\left(
U^{-1}\right) ^{\dagger }U^{-1}\tilde{H}UU^{\dagger }=\eta \tilde{H}\eta
^{-1}
\end{equation}
where $\eta =\left( U^{-1}\right) ^{\dagger }U^{-1}$ is Hermitian and
positive definite. We say that $\tilde{H}$ is $\eta $-pseudo-Hermitian\cite
{M02a,M02b,M02c} and (\ref{eq:metric}) becomes
\begin{equation}
\left\langle \varphi _{m}\right| \eta \left| \varphi _{n}\right\rangle
=\delta _{mn}.
\end{equation}

If $A$ and $B$ are two linear operators then
\begin{equation}
\lbrack \tilde{A},\tilde{B}]=U[A,B]U^{-1}.  \label{eq:[A',B']}
\end{equation}
In particular, the commutator $[x,p]=iI$ between the coordinate $x$ and
momentum $p$ is conserved
\begin{equation}
\lbrack \tilde{x},\tilde{p}]=iI.  \label{eq:[x',p']}
\end{equation}

\textit{Summarizing}: a non-Hermitian operator $\tilde{H}$ that is similar
or equivalent to an Hermitian one $H$ is pseudo Hermitian. In addition to
it, both operators are isospectral. When the similarity transformation is
unitary ($U^{-1}=U^{\dagger }$) it conserves the norm ($\left\langle \varphi
_{m}\right. \left| \varphi _{n}\right\rangle $ $=\delta _{mn}$), $\eta =I$
and $\tilde{H}$ is obviously Hermitian.

The results developed above are not new since they are contained in
Mostafazadeh's papers\cite{M02a,M02b,M02c}. We simply derived them here from
the point of view of a similarity transformation in order to connect them
with the papers of Ahmed\cite{A02} and Rath and Mallick\cite{RM15} in a
clearer way.

\section{Gauge-like transformation}

\label{sec:gauge}

The gauge-like transformation for one-dimensional operators
\begin{equation}
H=\frac{1}{2}p^{2}+V(x),  \label{eq:H(p,x)}
\end{equation}
discussed by Ahmed\cite{A02} is a particular case of the similarity
transformation outlined in Section~\ref{sec:similarity}. If we choose
\begin{equation}
U=e^{u(x)},  \label{eq:e^u(x)}
\end{equation}
then\cite{FC96}
\begin{equation}
\tilde{p}=UpU^{-1}=p+[u,p]=p+iu^{\prime },\;\tilde{x}=x,  \label{eq:p',x'}
\end{equation}
and
\begin{equation}
\tilde{H}=\frac{1}{2}(p+iu^{\prime })^{2}+V(x).
\end{equation}
Therefore, $H$ and $\tilde{H}$ are isospectral as discussed in Section~\ref
{sec:similarity}.

The transformation of the non-Hermitian operator
\begin{equation}
H_{\beta }=\frac{1}{2}[p+i\beta \nu (x)]^{2}+V(x),
\end{equation}
yields
\begin{equation}
\tilde{H}_{\beta }=\frac{1}{2}[p+i\beta \nu (x)+iu^{\prime }(x)]^{2}+V(x).
\end{equation}
If $\nu (x)$ is real and
\begin{equation}
u^{\prime }(x)=-2\beta \nu (x),
\end{equation}
then
\begin{equation}
\tilde{H}_{\beta }=H_{\beta }^{\dagger } .  \label{eq:H_beta_pseudo}
\end{equation}
Since $u(x)$ is real then $U$ is Hermitian and positive definite; therefore $%
H_{\beta }$ is $U$-pseudo-Hermitian.

In particular, Ahmed chose $\nu (x)=x$ and $V(x)=\left( \alpha ^{2}+\beta
^{2}\right) x^{2}/2$ so that
\begin{equation}
H_{\beta }=\frac{1}{2}(p+i\beta x)^{2}+\frac{1}{2}\left( \alpha ^{2}+\beta
^{2}\right) x^{2},
\end{equation}
and $u(x)=u_{1}(x)=-\beta x^{2}$ leads to equation (\ref{eq:H_beta_pseudo}).
Note that if $u_{2}(x)=-\beta x^{2}/2$ then
\begin{equation}
e^{u_{2}}H_{\beta }e^{-u_{2}}=\frac{1}{2}p^{2}+\frac{1}{2}\left( \alpha
^{2}+\beta ^{2}\right) x^{2}=H_{SHO}
\end{equation}
from which we conclude that $H_{\beta }$ and the simple harmonic oscillator $%
H_{SHO}$ are isospectral. In this case the eigenfunctions $\varphi _{n}(x)$
of the former operator are square integrable provided $\alpha \neq 0$\cite
{A02}. These results are particular cases of those derived in Section~\ref
{sec:similarity} (note that $e^{u_{2}}\left(e^{u_{2}}\right)^%
\dagger=e^{u_{1}}$).

Ahmed\cite{A02} also discussed the particular case $\beta =i\gamma $, $%
\gamma $ real, that leads to the Hermitian operator
\begin{equation}
H_{\gamma }=\frac{1}{2}(p-\gamma x)^{2}+\frac{1}{2}\left( \alpha ^{2}-\gamma
^{2}\right) x^{2},
\end{equation}
and draw two curious conclusions. He stated that ``Remarkably, the usual
connection between the nodal structure with the quantum number $n$ does not
hold any more. Even the ground state may have nodes for some values of $%
\gamma $.'' Since $|\varphi _{n}(x)|=|\psi _{n}(x)|$ it is obvious that both
functions have the same number of nodes; in particular, the ground state $%
\varphi _{0}(x)$ is nodeless in the interval $(-\infty ,\infty )$ as
expected. He also said that ``Eigenvalues (18) possess an interesting
feature of becoming complex (conjugate) at the cost of eigenfunction (19)
being delocalized as it would not vanish at $x=\pm \infty $. This
interesting phase-transition of eigenvalues from real to complex takes place
when $\gamma >\gamma _{critical}$ ($=\alpha $).'' It is obvious that this
\textit{interesting phase transition} is due to the force constant chosen
for $H_{SHO}$ and has nothing to do with the transformation of one
oscillator into the other. To see this point more clearly just choose
\begin{equation}
H_{\gamma }=\frac{1}{2}(p-\gamma x)^{2}+\frac{1}{2}kx^{2},
\end{equation}
and the phase transition does not take place for any value of $\gamma $ if $%
k>0$.

\section{Transformation of coordinate and momentum}

\label{sec:trans_x_p}

Recently, Rath and Mallick\cite{RM15} proposed the following generalization
of the gauge-like transformation:
\begin{equation}
x\rightarrow \tilde{x}=\frac{1}{\sqrt{1+\alpha \beta }}\left( x+i\alpha
p\right) ,\;p\rightarrow \tilde{p}=\frac{1}{\sqrt{1+\alpha \beta }}\left(
p+i\beta x\right) ,  \label{eq:x',p'_RM}
\end{equation}
that converts
\begin{equation}
H_{HO}=\frac{1}{2}\left( p^{2}+x^{2}\right)  \label{eq:H_HO}
\end{equation}
into the non-Hermitian operator
\begin{equation}
H=\frac{1}{2(1+\alpha \beta )}\left[ \left( p+i\beta x\right) ^{2}+\left(
x+i\alpha p\right) ^{2}\right] .  \label{eq:H_RM}
\end{equation}
By means of a non-rigorous procedure based on second quantization, an
adjustable frequency and a truncated perturbation expansion they conjectured
that the eigenvalues of $H$ appeared to be exactly those of $H_{HO}$.

This conclusion follows straightforwardly from the similarity transformation
\begin{equation}
H=UH_{HO}U^{-1},
\end{equation}
where $U$ is given by
\begin{equation}
UxU^{-1}=\tilde{x},\;UpU^{-1}=\tilde{p}  \label{eq:x',p'_RM2}
\end{equation}
According to the results of Section~\ref{sec:similarity} both operators are
isospectral with eigenvalues
\begin{equation}
E_{n}=n+\frac{1}{2},\;n=0,1,\ldots ,
\end{equation}
and $H$ is $\eta $-pseudo-Hermitian.

It only remains to determine whether the eigenfunctions of $H$ are square
integrable. To this end we resort to the construction of the eigenvectors of
$H_{HO}$ in second-quantization form\cite{FC96}:
\begin{equation}
a\left| \psi _{0}\right\rangle =0,\;\left| \psi _{n}\right\rangle =\frac{1}{%
\sqrt{n!}}\left( a^{\dagger }\right) ^{n}\left| \psi _{0}\right\rangle ,
\label{eq:second_quant}
\end{equation}
where
\begin{equation}
a=\frac{1}{\sqrt{2}}\left( x+ip\right) ,\;a^{\dagger }=\frac{1}{\sqrt{2}}%
\left( x-ip\right) .
\end{equation}
It follows from equations (\ref{eq:x',p'_RM}), (\ref{eq:x',p'_RM2}) and (\ref
{eq:second_quant}) that
\begin{equation}
\tilde{a}\left| \varphi _{0}\right\rangle =0,\;\left| \varphi
_{n}\right\rangle =\frac{1}{\sqrt{n!}}\left( \tilde{a}^{\dagger }\right)
^{n}\left| \varphi _{0}\right\rangle .
\end{equation}
Since
\begin{equation}
\tilde{a}=\frac{1}{\sqrt{2(1+\alpha \beta )}}\left[ (1-\beta )x+i(1+\alpha
)p\right] ,
\end{equation}
then the ground state $\varphi _{0}(x)$ is a solution of the first-order
differential equation
\begin{equation}
\varphi _{0}^{\prime }(x)=-\frac{1-\beta }{1+\alpha }\varphi _{0}(x),
\end{equation}
that leads to
\begin{equation}
\varphi _{0}(x)=\left[ \frac{1-\beta }{\pi (1+\alpha )}\right] ^{1/4}\exp
\left[ -\frac{1-\beta }{2(1+\alpha )}x^{2}\right] .
\end{equation}
We appreciate that $\varphi _{0}(x)$ is square integrable (and,
consequently, also all the other eigenfunctions $\varphi _{n}(x)$) provided
that $\beta <1$ and $\alpha >-1$. The square-integrability of the
eigenfunctions was not discussed by Rath and Mallick\cite{RM15} in spite of
the fact that the conditions just given appear explicitly in the zero and
pole of their chosen frequency $\omega$ for Case II.

The operator that carries out the transformation (\ref{eq:x',p'_RM}) is of
the form\cite{FC96}
\begin{equation}
U=\exp \left( ax^{2}+bp^{2}\right) ,
\end{equation}
where
\begin{eqnarray}
i\alpha &=&\frac{\left( e^{2\sqrt{-ab}}-1\right) \sqrt{-ab}}{a\left( e^{2%
\sqrt{-ab}}+1\right) }  \nonumber \\
i\beta &=&\frac{\left( 1-e^{2\sqrt{-ab}}\right) \sqrt{-ab}}{b\left( e^{2%
\sqrt{-ab}}+1\right) },
\end{eqnarray}
that leads to $\alpha /\beta =-b/a$.

\section{A more general coordinate-momentum transformation}

\label{sec:gen_x,p_transf}

A more general similarity transformation is given by\cite{FC96}
\begin{eqnarray}
\tilde{x} &=&UxU^{-1}=U_{11}x+U_{12}p  \nonumber \\
\tilde{p} &=&UpU^{-1}=U_{21}x+U_{22}p,  \label{eq:x',p',Uij}
\end{eqnarray}
where
\begin{equation}
U_{11}U_{22}-U_{21}U_{12}=1,  \label{eq:det(U)=1}
\end{equation}
follows from the condition $[\tilde{x},\tilde{p}]=iI$. Since the matrix
elements $U_{ij}$ may be complex numbers the transformation depends on $8$
parameters that should satisfy two equations; therefore, there are only $6$
independent parameters and the transformation is given by an exponential
operator of the form\cite{FC96}
\begin{equation}
U=\exp \left[ \frac{a}{2}x^{2}+\frac{c}{2}(xp+px)+\frac{b}{2}p^{2}\right] ,
\label{eq:U(a,b,c)}
\end{equation}
where $a$, $b$ and $c$ are complex numbers.

The application of this similarity transformation to the harmonic oscillator
$H_{HO}$ (\ref{eq:H_HO}) yields the operator
\begin{eqnarray}
\tilde{H} &=&UH_{HO}U^{-1}  \nonumber \\
&=&\frac{1}{2}\left[ \left( U_{22}^{2}+U_{12}^{2}\right) p^{2}+\left(
U_{11}^{2}+U_{21}^{2}\right) x^{2}+\left( U_{21}U_{22}+U_{11}U_{12}\right)
\left( xp+px\right) \right] .  \nonumber \\
&&  \label{eq:H'(Uij)}
\end{eqnarray}
By means of well known operator formulas\cite{FC96} it is not difficult to
prove that
\begin{eqnarray}
U_{11} &=&\cosh (\theta )-\frac{c}{\theta }\sinh (\theta )  \nonumber \\
U_{12} &=&-\frac{b}{\theta }\sinh (\theta )  \nonumber \\
U_{21} &=&\frac{a}{\theta }\sinh (\theta )  \nonumber \\
U_{22} &=&\cosh (\theta )+\frac{c}{\theta }\sinh (\theta )  \nonumber \\
\theta &=&\sqrt{c^{2}-ab}\;.
\end{eqnarray}
In general, any operator of the form (\ref{eq:H'(Uij)}) with matrix elements
$U_{ij}$ that satisfy the condition (\ref{eq:det(U)=1}) is equivalent (and
therefore isospectral) to the harmonic oscillator (\ref{eq:H_HO}). It is
always $\eta $-pseudo-Hermitian and under certain conditions it may also be
Hermitian or PT-symmetric. For example, if $U_{22}^{2}+U_{12}^{2}$ and $%
U_{11}^{2}+U_{21}^{2}$ are both real and $\left(
U_{21}U_{22}+U_{11}U_{12}\right) $ purely imaginary, then $\tilde{H}$ is
PT-symmetric. The choice $U_{11}=U_{22}=1$,$\;U_{12}=0$, and $U_{21}=i\beta $
yields one of the examples given by Ahmed\cite{A02}. On he other hand, when $%
U_{11}=U_{22}=1/\sqrt{1+\alpha \beta }$, $U_{12}=i\alpha /\sqrt{1+\alpha
\beta }$, and $U_{21}=i\beta /\sqrt{1+\alpha \beta }$ we obtain the model
proposed by Rath and Mallick\cite{RM15}. Obviously, when the coefficients of
$p^{2}$, $x^{2}$ and $xp+px$ are real $\tilde{H}$ is Hermitian.

Arguing as in Section~\ref{sec:trans_x_p} we conclude that the
eigenfunctions $\varphi _{n}(x)$ of $\tilde{H}$ are square integrable
provided that
\begin{equation}
\Re \frac{U_{11}+iU_{21}}{U_{22}-iU_{12}}>0
\end{equation}

\section{Positive and negative eigenvalues}

\label{sec:pos_neg_eigen}

By a suitable choice of the adjustable frequency Rath\cite{R15} managed to
obtain negative harmonic-oscillator-like eigenvalues. However, the author
did not consider the square integrability of the eigenfunctions with
sufficient detail. In order to analyze this aspect of the problem we resort
to a different approach.

Consider the non-Hermitian Hamiltonian
\begin{equation}
H=h_{11}p^{2}+ih_{12}(xp+px)+h_{22}x^{2},  \label{eq:H(hij)}
\end{equation}
where $[x,p]=i$ and the coefficients $h_{ij}$ are real. In order to obtain
its spectrum we express the coordinate and momentum operators in terms of
the creation $a^{\dagger }$ and annihilation $a$ operators as
\begin{equation}
x=\frac{1}{\sqrt{2\omega }}\left( a+a^{\dagger }\right) ,\;p=i\sqrt{\frac{%
\omega }{2}}\left( a^{\dagger }-a\right) ,  \label{eq:(x,p)(a,a+)}
\end{equation}
where $[a,a^{\dagger }]=1$. The Hamiltonian operator (\ref{eq:H(hij)}) then
becomes
\begin{eqnarray}
H &=&\left( \frac{h_{11}\omega }{2}+\frac{h_{22}}{2\omega }\right) \left(
2a^{\dagger }a+1\right)  \nonumber \\
&&+\left( -\frac{h_{11}\omega }{2}+h_{12}+\frac{h_{22}}{2\omega }\right)
a^{2}  \nonumber \\
&&+\left( -\frac{h_{11}\omega }{2}-h_{12}+\frac{h_{22}}{2\omega }\right)
\left( a^{\dagger }\right) ^{2}.  \label{eq:H(a,a+)}
\end{eqnarray}

We expand every eigenvector $\left| \psi \right\rangle $ of $H$ in the basis
set of eigenvectors $\{\left| n\right\rangle ,\;n=0,1,\ldots \}$ of the
occupation number operator $a^{\dagger }a$
\begin{equation}
\left| \psi \right\rangle =\sum_{n=0}^{\infty }d_{n}\left| n\right\rangle
\label{eq:|psi>=dn|n>}
\end{equation}
that satisfy
\begin{equation}
a\left| n\right\rangle =\sqrt{n}\left| n-1\right\rangle ,\;a^{\dagger
}\left| n\right\rangle =\sqrt{n+1}\left| n+1\right\rangle .
\label{eq:(a,a+)|n>}
\end{equation}
It follows from (\ref{eq:H(a,a+)}) and (\ref{eq:(a,a+)|n>}) that
\begin{equation}
H\left| n\right\rangle =A_{n}(\omega )\left| n-2\right\rangle +B_{n}(\omega
)\left| n\right\rangle +C_{n}(\omega )\left| n+2\right\rangle ,
\label{eq:H|n>}
\end{equation}
where
\begin{eqnarray}
A_{n}(\omega ) &=&\left( -\frac{h_{11}\omega }{2}+h_{12}+\frac{h_{22}}{%
2\omega }\right) \sqrt{n(n-1)}  \nonumber \\
B_{n}(\omega ) &=&\left( \frac{h_{11}\omega }{2}+\frac{h_{22}}{2\omega }%
\right) \left( 2n+1\right)  \nonumber \\
C_{n}(\omega ) &=&\left( -\frac{h_{11}\omega }{2}-h_{12}+\frac{h_{22}}{%
2\omega }\right) \sqrt{(n+1)(n+2)}.  \label{eq:An,Bn,Cn}
\end{eqnarray}

It follows from $H\left| \psi \right\rangle =E\left| \psi \right\rangle $
and equation (\ref{eq:H|n>}) that
\begin{equation}
A_{n+2}d_{n+2}+\left( B_{n}-E\right) d_{n}+C_{n-2}d_{n-2}=0.
\label{eq:rec_rel_An,Bn,Cn}
\end{equation}

Note that $C_{n}(\omega )=0$ for all $n$ if
\begin{equation}
\omega =\left\{
\begin{array}{c}
\omega _{+}=\frac{\sqrt{h_{11}h_{22}+h_{12}^{2}}-h_{12}}{h_{11}}>0 \\
\omega _{-}=-\frac{\sqrt{h_{11}h_{22}+h_{12}^{2}}+h_{12}}{h_{11}}<0
\end{array}
\right. .  \label{eq:omega=omega_(+-)}
\end{equation}
For either of these values of $\omega $ we have
\begin{equation}
d_{n+2}=\frac{E-B_{n}}{A_{n+2}}d_{n},  \label{eq:d_(n+2)=}
\end{equation}
so that
\begin{equation}
\left| \psi _{k,s}\right\rangle =\sum_{j=0}^{k}d_{2j+s}\left|
2j+s\right\rangle ,\;E_{k,s}=B_{2k+s},  \label{eq:|psi>_(2k+s)}
\end{equation}
where $k=0,1,\ldots $ and $s=0$ or $s=1$ give us the even or odd
states, respectively. It is worth noting that the eigenvectors of
$H$ are not exactly those of the occupation number operator,
except when $k=0$. Rath\cite {R15}, on the other hand, appears to
suggest that both $H$ and $a^{\dagger }a $ have a common set of
eigenvectors in spite of the fact that these operators do not
commute.

The ground state eigenfunction $\psi _{0}(x)=\left\langle x\right| \left.
\psi _{0}\right\rangle $ obtained from $\left\langle x\right| a\left| \psi
_{0}\right\rangle =0$ is square integrable when $\omega >0$ as follows from
\begin{equation}
\psi _{0}(x)=\frac{|\omega |^{1/4}}{\pi ^{1/4}}\exp \left( -\omega
x^{2}/2\right) .  \label{eq:psi_0(x)}
\end{equation}
Therefore, for $\omega =\omega _{+}$ we have
\begin{equation}
E_{n}\left( \omega _{+}\right) =\sqrt{h_{11}h_{22}+h_{12}^{2}}\left(
2n+1\right) ,  \label{eq:En}
\end{equation}
where $n=2k+s$ takes into account the even and odd states simultaneously. On
the other hand, when $\omega =\omega _{-}$
\begin{equation}
E_{n}\left( \omega _{-}\right) =-\sqrt{h_{11}h_{22}+h_{12}^{2}}\left(
2n+1\right) ,
\end{equation}
and the eigenfunctions $\psi _{n}(x)=\left\langle x\right| \left. \psi
_{n}\right\rangle $ are square integrable along the imaginary axis $ix$.

The three-parameter Hamiltonian (\ref{eq:H(hij)}) is obviously more general
than the two-parameter one discussed by Rath and Malick\cite{RM15} and Rath%
\cite{R15} where
\begin{equation}
h_{11}=\frac{1-\lambda ^{2}}{2(1+\lambda \beta )},\;h_{12}=\frac{\lambda
+\beta }{2(1+\lambda \beta )},\;h_{22}=\frac{1-\beta ^{2}}{2(1+\lambda \beta
)}.  \label{eq:hij(alpha,beta)}
\end{equation}
Note that in this particular case $h_{11}h_{22}+h_{12}^{2}=\frac{1}{4}$ and
\begin{eqnarray}
\omega _{1} &=&\frac{1-\beta }{1+\lambda }  \nonumber \\
\omega _{2} &=&\frac{1+\beta }{\lambda -1}.  \label{eq:omega+-(beta,lambda)}
\end{eqnarray}

\section{Conclusions}

\label{sec:conclusions}

The purpose of this paper is to show that the results of Ahmed\cite{A02} and
Rath and Mallick\cite{RM15} can be straightforwardly derived and proved by
suitable similarity transformations. In the former case there is no need of
discussing the reality of the operator and its eigenfunctions or the
orthogonality conditions. In fact, the proposition enunciated by the author
does not explain the situation. Once we prove that a non-Hermitian operator
is similar to an Hermitian one the reality of the spectrum of the former is
certainly proved. Of course, caution must be exercised with respect to the
square-integrability of its eigenfunctions.

With respect to the latter paper\cite{RM15} the similarity transformation is
a much more rigorous and straightforward way of proving that the
non-Hermitian operator is isospectral with the harmonic oscillator. The
results of both papers are merely particular cases of the general
expressions derived by Mostafazadeh\cite{M02a,M02b,M02c} and also of the
equations derived in Section~\ref{sec:similarity}.

Equation (\ref{eq:H'(Uij)}) with the restriction (\ref{eq:det(U)=1}) enables
us to construct a family of non-Hermitian operators with real spectrum. If
necessary we can enlarge the number of cases by choosing $H_{HO}=p^{2}+kx^{2}
$, $k>0$, instead of the operator (\ref{eq:H_HO}) thus having one more
independent parameter at our disposal.

We have also shown how to obtain the eigenvalues and eigenvectors of a more
general three-parameter oscillator by a judicious modification of the
procedure proposed by Rath and Mallick\cite{RM15} and Rath\cite{R15}.
Present approach is completely rigorous (unlike the perturbation approach%
\cite{RM15}) and reveals that the eigenvectors of the non-Hermitian operator
are not exactly those of the occupation number operator (as suggested by Rath%
\cite{R15}) but linear combinations of them.

\end{document}